\documentclass[12pt,preprint]{aastex}
\slugcomment{Accepted by ApJ}
\shorttitle{The periods in AO 0235+164}
\shortauthors{J.H. Fan et al.}

\begin{document}
\title{The Variability and Period Analysis for the BL Lac AO 0235+164}
\author{J.H. Fan\altaffilmark{1,2}{\thanks{email:fjh@gzhu.edu.cn}}, O. Kurtanidze\altaffilmark{3,4},  Y. Liu\altaffilmark{1,2},  X. Liu\altaffilmark{5}, J.H. Yang\altaffilmark{6}, G.M. Richter
\altaffilmark{7}, M.G. Nikolashvili\altaffilmark{3}, S. O. Kurtanidze\altaffilmark{3}, H. T. Wang\altaffilmark{8}, M. Sasada\altaffilmark{9}, A. Y. Zhou\altaffilmark{10}, C. Lin\altaffilmark{1,2}, Y. H. Yuan\altaffilmark{1,2}, Y. T. Zhang\altaffilmark{1,2}, D. Constantin\altaffilmark{1,2}}
\altaffiltext{1}{Center for Astrophysics, Guangzhou University,
Guangzhou 510006, China}
 \altaffiltext{2}{Astronomy Science and Technology Research Laboratory of
   Department of Education of Guangdong Province, Guangzhou 510006,
   China}
 \altaffiltext{3}{Abastumani Observatory, Mt. Kanobili, 0301 Abastumani, Georgia}
 \altaffiltext{4}{Engelhardt Astronomical Observatory, Kazan Federal University, Tatarstan, Russia}
 \altaffiltext{5}{Xinjiang Astronomical Observatory, Chinese Academy of Sciences, Urumqi 830011, China}
 \altaffiltext{6}{Dept of Physics and Electronics Science, Hunan
University of Arts and Science, Changde, 415000, China}
  \altaffiltext{7}{Astrophysikalisches Institut Potsdam, An der Sternwarte 16, 14482 Potsdam, Germany}
 \altaffiltext{8}{Faculty of Physics and Electronic Information, LangFang Teacher's College, China}
 \altaffiltext{9}{Department of Astronomy, Boston University, USA}
 \altaffiltext{10}{National Astronomical Observatory, Chinese Academy of Sciences, Beijing, China}

\begin{abstract}
Variability is one of the extreme observational properties of BL Lacertae objects.  AO 0235+164 is a well studied BL Lac through the whole electro-magnetic wavebands. In the present work, we show its optical R band photometric observations carried out during the period of Nov, 2006 to Dec. 2012  using    the  Ap6E CCD camera attached to the primary focus of
the $\rm 70-cm$ meniscus telescope at Abastumani   Observatory, Georgia. It shows a large variation of $\Delta R$ = 4.88 mag (14.19 - 19.07 mag) and  a short time scale of $\Delta T_v$ = 73.5 min  during our monitoring period.  During the period of Dec. 2006 to Nov. 2009, we made radio observations of the source using the 25-m radio telescope at Xinjiang Astronomical Observatory.  When  a discrete correlation function (DCF) is adopted  to the optical and radio observations, we found that the optical variation leads the radio variation
by 23.2$\pm$12.9 days.
\end{abstract}


\section{Introduction}

Blazars as a very extreme subclass of active galactic nuclei(AGNs)  show special observation properties, such as luminous emissions, rapid and high amplitude variability, high and variable polarization, strong and variable $\gamma$-ray emissions, strong emission line feature or no-emission line features at all, or  superluminal motions etc. Blazars have two subclasses: BL Lacertae objects (BLs) and  flat spectrum radio quasars (FSRQs). The major difference for the two subclasses is  their difference in emission line features with  FSRQs showing strong emission line features while BLs showing very weak emission lines or no emission lines at all.  BLs can be divided into radio selected BLs (RBLs) and X-ray selected BLs (XBLs) from survey, or low frequency-peaked BL Lacertae objects (LBLs, ${\rm log}\nu_{\rm p} < 15$ Hz) and high frequency-peaked BL Lacertae objects (HBL, ${\rm log} \nu_{\rm p} > 15$ Hz) from the synchrotron peak frequency in their spectral energy distribution (SED)
\citep{padovani1995,Urry1995}.
To avoid confusion, Abdo et al. (2010a) extended the definition to all types of non-thermal dominated AGNs using new
acronyms as low synchrotron peaked blazars-LSP, intermediate synchrotron peaked blazars-ISP, and high synchrotron peaked blazars-HSP  with
LSP showing their peak synchrotron powers in  far-IR or IR band (${\rm log}\nu_{\rm p} < 14$ Hz);
ISP  showing their peak synchrotron emissions in the frequency range of
   ${\rm log}\nu_{\rm p} = 14 \sim 15$ Hz; and
HSP showing their peak synchrotron powers at frequency of ${\rm log}\nu_{\rm p} > 15$ Hz.
Very recently, we calculated the SEDs using ${\rm log}(\nu F_{\nu}) = P_1({\rm log}\nu - P_2)^2 + P_3$ for a sample of 1425 Fermi blazars. Synchrotron peak frequency ($\rm log \nu_p$), spectral curvature ($\rm P_1$),  peak flux ($\rm \nu_p F_{\nu_p}$), and integrated flux ($\rm \nu F_{\nu}$)  are successfully obtained for 1392 blazars.  The "Bayesian classification" is adopted to log$\nu_{\rm p}$ in the rest frame for 999 blazars with available redshift and the results show that  3 components are enough to fit the log$\nu_{\rm p}$
 distribution. Therefore, we proposed
LSP with ${\rm log}\nu_{\rm p} < 14$ Hz;
ISP with ${\rm log}\nu_{\rm p} = 14 \sim 15.3$ Hz;
HSP with ${\rm log}\nu_{\rm p} > 15.3$ Hz \citep{Fan2016a}.

One of the most important results of the Fermi/LAT is the discovery of blazars,
which emit most of their bolometric  luminosity in the high energy
range $\gamma$-rays  (0.1 $\sim$ 100 GeV) \citep{1FGL,Ackermann2012,2FGL,3LAC}.
The $\gamma$-rays are found to be strongly beamed \citep{Fan2014a,
 Hovatta2009,Savolainen2010,
Ackermann2011,
 Giroletti2012,
 Giovannini2014}.

Variability is one of the typical observation properties of blazars, which
show variabilities at almost the whole electromagnetic
 wavebands (Ackermann et al. 2012 and reference therein).
 The variations have been found to be over  time scales from less than one hour
to as long as years Fan (2005), who divided the time scales into three
classes:
 micro-variability (intra-day variability, or IDV) with time scale, $\Delta T$ being less than one day,
 short-term variation (STV) with $\Delta T$ being one day to several months, and
 long-term variation (LTV) with $\Delta T$ being longer than one year.
 From observations, we can see that the short-term variations are non-periodic
  while the long-term variation in some cases is quasi-periodic as discussed in literatures
  \citep{
  Jurkevich1971,
  Sillanpaa1988,
  Fan1998,
  Fan2002,
  Cia04,Wu2006, Ciprini07,
  Fan2007, Valtonen2008,
  Rani2010,Wii11,
  QianT2004,
  Gupta2014,
  Gaur2015a,
  Gaur2015b}.

Photometric monitoring programme can also provide opportunity for people to investigate  possible variability periods.  The long-term variability period has been explained by various mechanisms \citep{Cia04}, such as shocks in jets, changes in the direction of forward beaming, and precession in a binary black-hole system \citep{Sillanpaa1988, CK92, RM00,  Valtonen2008}. It has been claimed that the possible periodicity in the historical light curves also shows helical trajectories in their VLBI radio components \citep{VR99}.

BL Lac AO 0235+164 ($02^h38^m38.9^s\,\, +16^d36^m59^s$(2000.0), A$_{R}$ = 0.173), located at $z_{EM}$ = 0.94 \citep{Cohen1987},  is a well studied object. Its earlier optical spectroscopy revealed two absorption line systems, one
at $z_{abs} = 0.524$  and the other one at $z_{abs} = 0.852$ discovered by Burbidge et al. (1976) and by Rieke et al. (1976).
It is observed from radio to X-ray bands, and even high energetic $\gamma$-ray regions.
It shows variability timescales from less than one hour   to several years
\citep{Webb2000,
  Romero1997, Romero2000,
  Fan2002,
  Peng2004,Hagen2008,
  Ackermann2012,
  Wang2014,
  Volvach2015}.
Its historic  optical variation is over 5 magnitudes
\citep{Rieke1976,Webb2000}. In our previous paper, variations in the UBVRI bands are
$\Delta U = 4.26$,
$\Delta B = 5.47 $,
$\Delta V = 4.74$,
$\Delta R = 4.18 $, and
$\Delta I = 3.85 \rm \,mag.$ \citep{FanL2000}.

AO 0235+164 has been the target of several WEBT campaigns
\citep{Raiteri2001,Raiteri2005,Raiteri2006,Raiteri2008} and one of the GASP sources
\citep{Ackermann2012}. It shows a bluer-when-brighter (BWB) trend when it was in  bright flares but no clear BWB trend when it was in faint low amplitude flares as claimed by Sasada et al. (2011) (see also Sasada 2012).
It  is also a highly polarized BL Lac object. Sasada (2012) found that the polarization varied from 0 to $\sim$ 30\%, and the high polarizations correspond to bright outbursts.  Polarization variation  between 35\% and 13\% on nightly time scale is recently reported \citep{Larionov2015}.
The highest degree of polarization,
$ P\,= \,43.9 \%$  was reported by Impey et al. (1982), and the polarization value was renewed to be
$ P\, \sim \,50 \%$ in the paper \citep{Hagen2008}.
Raiteri et al. (2001)  analyzed about 25 years of observational data in optical and radio bands
during the period   from 1975 to 2000, and found  a quasi-periodicity of the main radio (and optical)
outbursts on a 5.7-year time scale. A period of 5.87$\pm$1.3 years was found in our pervious work
based on 16 years of optical observations \citep{Fan2002}, but $5.8\pm0.3$ years (based on 14.5 GHz light curve), $5.7\pm0.3$ years (based on 8.0 GHz light curve),  and $10.0\pm1.3$ years (based on 4.8 GHz light curve) are found in its radio bands \citep{Fan2007}. It  perhaps suggests existence of a binary black hole system at its center
\citep{Romero2003,Ostorero2004}. Very recent analysis based on the long term multiwavelength observations implies  that AO 0235+164 hosts a close binary supermassive black holes with similar masses of the order of 10$^{10} M_{\odot}$ \citep{Volvach2015}.  Variabilities on time scales of $\sim$ 17 days, $\sim$162 days, and $\sim$275 days were also reported \citep{Rani2009}. Short variability time scale is from 0.31 hrs in polarization variability to 6.15 hrs in V band photometry \citep{Hagen2008} while  extreme intra-night variability with amplitudes of $\sim 100\%$ over time scales of 24 hours, and changes of 0.5 magnitudes in both R and V bands  within a single night, and variations of 1.2 magnitudes from night to night were detected   \citep{Romero2000}.
 AO 0235+164 is one of the objects in our monitoring programme at Abastumani Observatory, Georgia
\citep{Kur07,  NK07, Fan2004,Fan2014b}.

The paper is arranged as follows. In section 2, we  describe the
observations and the data reduction; in section 3, give analysis results; and in section
4, give discussions and conclusions.

\section{Observations}
\subsection{Optical Observations and Data Reduction}

Abastumani Observatory is located at the top of the Mountain  Kanobili in the
South-Western part of Georgia. Mt. Kanobili is about 1,700 meters above the
sea level with a latitude of $41^{\circ}.8051$ and a longitude of
$42^{\circ}.8254$ respectively. The weather and seeing conditions are excellent
(about $1/3$ clear nights per year with seeing $\leq 1 \,{\rm arcsec}$). The
mean values of the night sky brightness are $B = 22.0$, $V = 21.2$, $R = 20.6$,
and $I = 19.8 \,{\rm magnitude}$.

All our observations were made using a 70 cm meniscus telescope (f/3), to
which a Peltier cooled ST-6 CCD imaging camera was attached to the
Newtonian focus from March 1997 to Sept 2006. Post October 2006 an Apogee Ap6E CCD camera (1024$\times$1024, 24$\times$24 micron,
quantum efficiency are 0.40 and 0.72 at 400 nm and 560 nm, respectively)
was attached to a primary focus. The readout, digitizing, downloading
time is 1 sec. We used only the  central portion 350$\times$350 square pixels (15$\times$15
square arcmin), while entire FOV is 40$\times$40 square arcmin.

All our observations are made using the filter $R_C$ passband \citep{KN99}. Our exposure times are 60  to 300 seconds.   The processing of image frames (bias correction, flat fielding, cosmic rays removal, etc.) and the photometry of the calibrated image frames are
carried out by the standard routines in Daophot II.  For Daophot II, the aperture was of a fixed diameter size  of 10 arc seconds. See our previous work (Fan et al. 2014b) for details.

For the comparison stars (S$_i$, $i= 1, 2, 3, N$) and the target ($O$), we calculate the differential magnitude, $O-S_i$,  and the corresponding uncertainties $\sigma_{O-S_i}$. For the comparison stars, we also calculate the differential magnitudes ($\Delta m_{\rm {ij}}$), and the corresponding uncertainties  ($\sigma_{\Delta m_{\rm {ij}}}$) of any two  comparison stars in the field. Here $\Delta m_{\rm {ij}} = m_{\rm i}-m_{\rm j}$, $m_{\rm i}$ and $m_{\rm j}$ are the magnitudes of comparison stars, S$_{\rm i}$ and S$_{\rm j}$ respectively. For the uncertainty calculation, we  take into account CCD chip parameters, sky background, source and comparison star counts (See Kurtanidze \& Nikolashvili, 2002). Finally, we choose the two stars which show the minimum deviation as our comparison stars, S$_{1}$ and S$_{2}$. The magnitude of the object can be determined using S$_{1}$ ( or S$_{2}$ ), the corresponding uncertainty of $\sigma_{O-S_1}$ ( or $\sigma_{O-S_2}$ ) is taken as the uncertainty of the observation.

 Romero et al. (1999) (see also Cellone et al. 2000
and Fan et al. 2001) introduced a variability parameter,  $C_{i} = {\frac{\sigma_{(O-S_i)}}{\sigma_{(S_1-S_2)}}}$, $i$ = 1 and 2, to check the reality of a variability.  Here, $\sigma_{(O-S_i)}$ is the deviation of the difference of the target object and the comparison star, $\sigma_{(S_1-S_2)}$ is the deviation of the two comparison stars. If $C ( = \frac{C_1 + C_2}{2})$, the average value of $C_1$ and $C_2$, is greater than 2.576, then the nominal confidence level of a variability detection is greater than 99\%.

In 2012, Gaur et al. adopted  the standard $F$-test discussed by de Diego (2010). The $F$-test
 is a properly distributed statistics. For two samples, one is the object differential
light curve measurements, the other is differential
light curve measurements of comparison star. If  their variances are $S_O^2$ and $S_C^2$ respectively, then
 $F={\frac{S_O^2}{S_C^2}}.$

For observations, the number of degrees of freedom for each sample, $\nu_O$
and $\nu_C$ are the same and equal to the number of measurements,
N, minus 1 ( $\nu$ = N-1). To investigate the reality of variation in a source, we compare the $F$ value with the
critical value, $F_{C(\nu_O,\nu_C)}(\alpha¦Á)$, where $\alpha$ is the significance
level set for the test. The smaller the $\alpha$, the
more improbable that the result is produced by chance.
If $F$ is greater than the critical value, the null hypothesis (no
variability) is discarded. We have performed F-tests at two
significance levels (1\% and 0.1\%) which correspond to
 2.6$\sigma$ and 3$\sigma$  detections respectively (See de Diego£¬ 2010£¬ and Gaur et al. 2012 for details).

In our previous paper, we proposed that a variability can be taken as real  if the variability is 3 times greater than the deviation, namely $\Delta m_{12}= m_{1} - m_{2} \ge 3 \sqrt{\sigma_1^2+\sigma_2^2}$, where $\sigma_1$ and $\sigma_2$ are the uncertainties corresponding to m$_{1}$ and m$_{2}$, the corresponding time interval is adopted as the time scale $\Delta$T = t$_{m2}$-t$_{m1}$ \citep{Fan2009}. The time scale can also be expressed as
$\Delta T_v = \Delta S/(dS/dt)$, here $S$ is the flux density.

The variability amplitude can be espressed as \citep{Heidt1996}:
\begin{equation}
A=100\times\sqrt{(m_{max}-m_{min})^{2}-2\sigma^{2}}(\%),
\label{amp}
\end{equation}
here $m_{max}$ and $m_{min}$ are the maximum and minimum magnitudes in the light
curves, $\sigma$ is the averaged measurement error of the observing run.

\begin{figure}
  \centering \includegraphics*[width=0.6\linewidth]{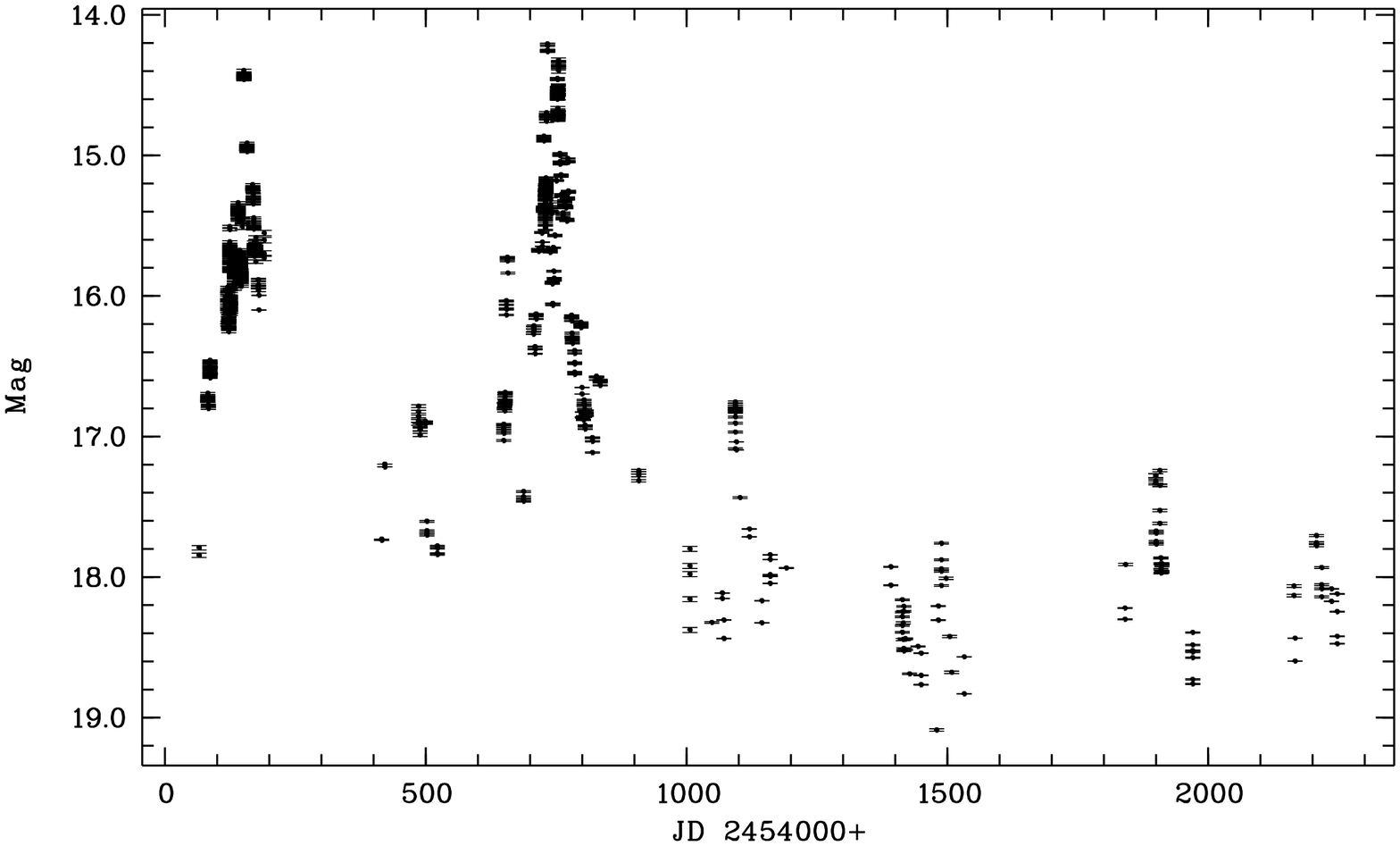}
  \caption{ Optical R photometry results of 0235+164 during the observing period of 2006 to 2012. }
\label{Fan-2016-ApJS-0235-LC-O}
\end{figure}

In this paper, the optical $R-$band observations were carried out from Nov. 2006 to Dec. 2012. The data obtained in  present paper have used the
standard stars  8, 9, 10, and 11 \citep{Kidger2001}. From our calculations, the comparison stars 9 and 11 are satisfied for the
minimum deviation,  $\sigma_{9-11}$ = 0.008, and are used as comparison stars in our
photometry determinations of AO 0235 + 164.
 The $R-$band  observations are listed in Table \ref{Fan-2016-ApJS-0235-obs-opt}, the light curve for AO 0235 + 164 is shown in Fig. \ref{Fan-2016-ApJS-0235-LC-O}.

\subsection{Radio Observations and Data Reduction}

The flux density monitoring observations of AO 0235+164 were carried out at more or less a monthly
sampling rate at 4.8 GHz from Dec. 2006 to Nov. 2009 with the
Urumqi 25-m radio telescope, with the central frequency of 4.8 GHz and bandwidth of 600 MHz. The typical system
temperature is 24~K in clear weather, and the antenna sensitivity
is $\sim$0.12 K/Jy.

\begin{figure}
  \centering \includegraphics*[width=0.6\linewidth]{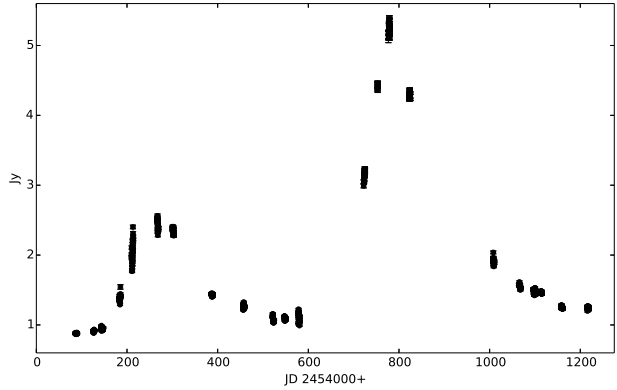}
  \caption{ Radio light curve of AO 0235+164 during 2006 to 2009. }
\label{Fan-2016-ApJS-0235-LC-Radio}
\end{figure}

The observations were performed in cross-scans mode, consisting of 8 sub-scans in azimuth and elevation over the source
position. After initial calibration of the raw data, the intensity profile of each sub-scan was fitted with a Gaussian function after subtracting a baseline, then the fitted scans were averaged in azimuth and elevation, respectively. After this  correction for residual pointing errors was obtained and the elevation and azimuth scans were averaged together. In the next step an antenna gain elevation correction was applied, including the correction for air mass. The antenna gain-elevation correction was derived from frequent observations of secondary calibrators observed during each observing run. These calibrators were further used to correct the data for systematic time-dependent effects. Finally, the raw amplitudes were converted to the absolute flux density using the average scale of the primary calibrators, e.g. 3C48, 3C286 \citep{Baars1977,Ott1994}.
The radio observations are listed in Table \ref{Fan-2016-ApJS-0235-obs-rad} and shown in Fig.  \ref{Fan-2016-ApJS-0235-LC-Radio}.

\section{Results and Analysis}

\subsection{Optical Results}

From Table \ref{Fan-2016-ApJS-0235-obs-opt} (or light curve in Fig. \ref{Fan-2016-ApJS-0235-LC-O}), we can see that AO 0235+164 is extreme variable. It shows a variation of $\Delta R\,\sim$\,\,4.88 mag from $R$ = 14.19 to $R$ = 19.0695 mag in the whole observing period. The light curve shows 6 clear peaks at
JD 2454151 (R = 14.3738 mag),
JD 2454754 (R = 14.3258 mag),
JD 2455093 (R = 16.8380 mag),
JD 2455488 (R = 17.7453 mag),
JD 2455907 (R = 17.2259 mag), and
JD 2456207 (R = 17.7349 mag). The first two peaks occurred at an interval of 603 days, and then the sources declined to R = 19.0695 at 2455479, next it brightened again to R = 17.7453 mag in 9 days, and finally  to   R = 17.2259 mag at JD 2455907.

\begin{figure}
  \centering \includegraphics*[width=1.0\linewidth]{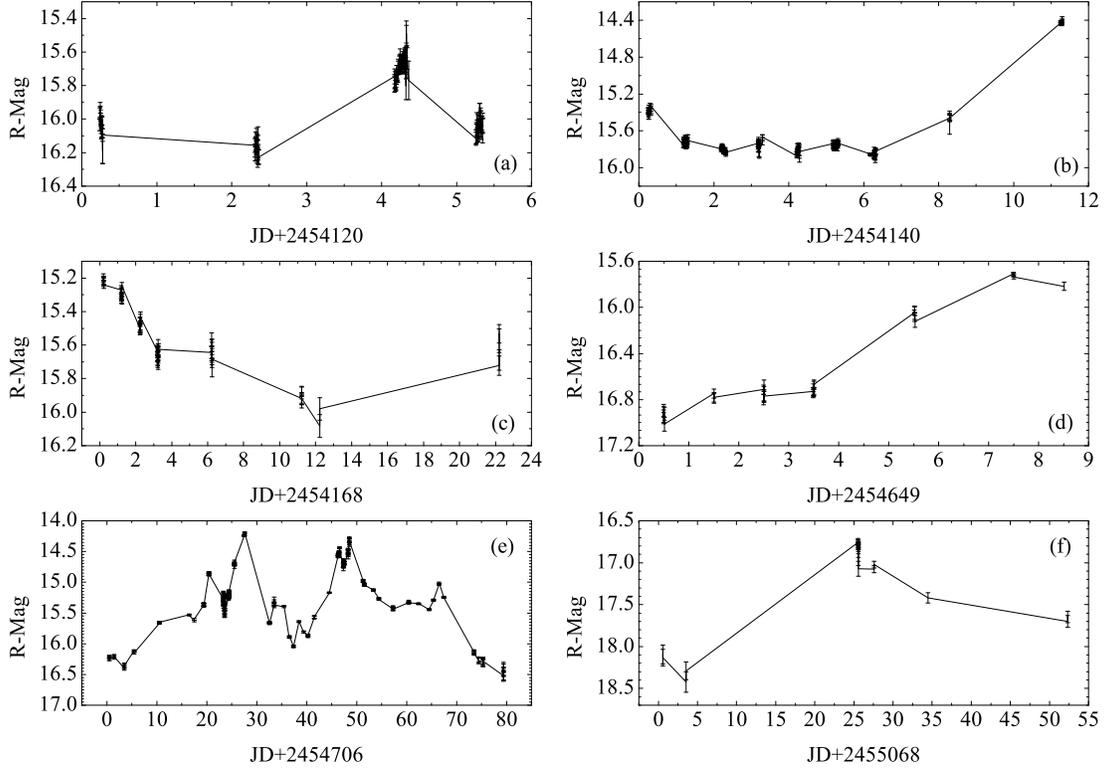}
  \caption{ Short-term variability of AO 0235+164.
  (a) during JD 2454120 to JD 2454125,
  (b) during JD 2454140 to JD 2454151,
  (c) during JD 2454168 to JD 2454190,
  (d) during JD 2454649 to JD 2454657,
  (e) during JD 2454706 to JD 2454785, and
  (f) period of JD 2455068 and 2455120.}
\label{Fan-2016-ApJS-0235-LC-STV}
\end{figure}

\begin{figure}
  \centering \includegraphics*[width=0.7\linewidth]{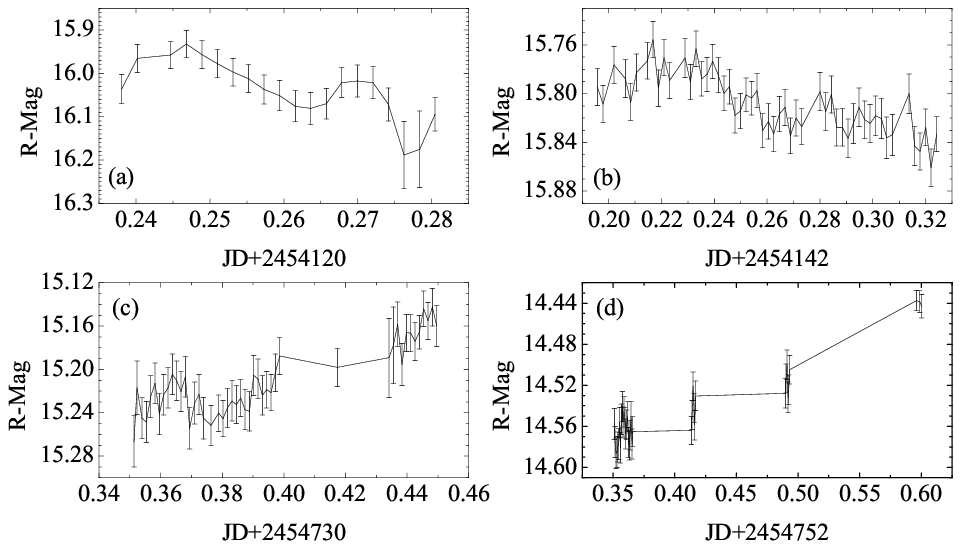}
  \caption{ Intraday variability of AO 0235+164,
  (a) at JD 2454120,
  (b) at JD 2454142,
  (c) at JD 2454730, and
  (e) at JD 2454752}
\label{Fan-2016-ApJS-0235-LC-IDV}
\end{figure}

From the works \citep{Heidt1996,Fan2009}, we can discuss   variability amplitude, $A$, and variability time scale ($\Delta T$).  During a 5-day observing period of JD 2454120 to JD 2454125, its brightness changed  from
$\langle R \rangle = 16.0370\,\pm 0.0659 $ at JD 2454120 to
  $\langle R \rangle = 16.1697\,\pm 0.0414 $ at JD 2454122, and to
 $\langle R \rangle = 15.6913\,\pm 0.0660 $ at JD 2454124, and afterwards  declined to
$\langle R \rangle = 16.0427\,\pm 0.0430 $ at JD 2454125. A brightening of $A = 73.35\%$ in two days (from JD 2454122 to JD 2454124) followed by a dimming of $A = 62.44\%$ within one  day, the corresponding variability parameter is $ C\, = £¨C_1 + C_2)/2 = (25.43+25.21)/2 = 25.32 $. The light
 curve is shown in Fig. \ref{Fan-2016-ApJS-0235-LC-STV}(a). The variation, and the corresponding $C$ values and the $F$ values are listed in Table \ref{Table:VI-STV}.
 At JD 2454120, its brightness decreased from R = 15.932 to R = 16.188 within 42 min, the corresponding variability
  amplitude, time scale, and variability parameter are $A = 25.0\%$, $\Delta T_v = 73.5$  min, and $C\,=\, 9.58$
  (see Fig. \ref{Fan-2016-ApJS-0235-LC-IDV}(a)). The time scale, variability, and the corresponding $C$ values and the $F$ values are listed in Table \ref{Table:VI-IDV}.

\begin{table}
\caption{Results of short-term  variability (STV) of AO 0235+164} \centering
\begin{tabular}{l|c|c|c|c}
\hline
Observation Time &  N  & C-Test         & F                                     & Variable  \\
                 &     & $C_1$, $C_2$, $C$    & $F_1$, $F_2$, $F_c(0.99)$, $F_c(0.999)$ &         \\
\hline
JD2454120-2454125	&	147	&	25.43	,	25.21	,	25.32	&	645.10	,	632.28	,	1.48	,	1.68	&	V 	\\
JD2454140-2454152	&	283	&	46.21	,	46.37	,	46.29	&	2132.98	,	2137.65	,	1.33	,	1.45	&	V 	\\
JD2454168-2454190	&	76	&	19.96	,	19.88	,	19.92	&	397.26	,	393.82	,	1.72	,	2.06	&	V 	\\
JD2454649-2454657	&	34	&	75.45	,	75.47	,	75.46	&	5670.62	,	5657.99	,	2.28	,	3.04	&	V 	\\
JD2454706-2454785	&	342	&	49.98	,	49.90	,	49.94	&	2475.78	,	2476.47	,	1.27	,	1.40	&	V 	\\
JD2455068-2455120	&	23	&	95.14	,	95.36	,	95.25	&	9032.52	,	9029.95	,	2.79	,	3.99	&	V 	\\
\hline
\end{tabular}
\label{Table:VI-STV}
\end{table}

\begin{table}
\caption{Results of intra-day variability (IDV) of AO 0235+164} \centering
\begin{tabular}{l|c|c|c|c|c|c}
\hline
Observation Time &  N  & C-Test         & F                                     & Variable & A(\%) & $\Delta T$ \\
                 &     & $C_1$, $C_2$, $C$    & $F_1$, $F_2$, $F_c(0.99)$, $F_c(0.999)$ &          &       & (min.)   \\
\hline
JD2454120	&	20	&	9.30	,	9.85	,	9.58	&	86.50	,	96.85	,	3.13	,	4.69	&	V 	&	25.0	&	73.5	\\
JD2454142	&	52	&	5.33	,	5.33	,	5.33	&	28.30	,	23.85	,	1.94	,	2.42	&	V 	&	10.0	&	252	\\
JD2454730	&	45	&	5.15	,	4.67	,	4.91	&	26.56	,	21.48	,	2.04	,	2.60	&	V 	&	10.87	&	144.5	\\
JD2454752	&	36	&	5.09	,	4.91	,	5.00	&	25.39	,	24.00	,	2.23	,	2.93	&	V 	&	10.74	&	499	\\
\hline
\end{tabular}
\label{Table:VI-IDV}
\end{table}

 During the period of JD 2454140 to JD 2454151, it decreased from R = 15.3382 to 15.8608 within 2 days, corresponding to $\Delta R \sim 0.528$ mag over 2 days and kept this state for 4 days, then brightened again to 15.4587 within 2 days and finally  to 14.3738 in the following 3 days, suggesting a brightening  of $\Delta R \sim 1.537$ mag within 5 days (see Table \ref{Table:VI-STV} and Fig. \ref{Fan-2016-ApJS-0235-LC-STV}(b)).
  The corresponding variability parameter during this period is $ C\, =\, 17.4$.  At
   JD 2454142, there is a variability with $A $ = 10\% within 152 min (corresponding to a  time scale of $\Delta \,T_v = 252\,\, \rm{min.}$), see
   Table \ref{Table:VI-IDV} and      Fig. \ref{Fan-2016-ApJS-0235-LC-IDV}(b)).

 During the period of JD 2454168 to JD 2454190, it decreased from R = 15.2222 to 15.7091 within 3 days,  and kept this state for  3 days, and then it decreased to R = 16.0837 (JD 2454180) within 4 days, corresponding to $\Delta R \sim 0.8173 $ mag over 12 days. After that it increased to R = 15.5333 at JD 2454190, corresponding to a variation of $\Delta R \sim 0.4204$ mag over 10 days. See
 Table \ref{Table:VI-STV} and  Fig. \ref{Fan-2016-ApJS-0235-LC-STV}(c).

 During the period of JD 2454649 to JD 2454657, its brightness increased from R = 17.0138 (JD 2454649) to R  = 15.7101 (JD 2454656), corresponding to  a variability of $\Delta R \sim 1.3037$ mag over 7 days. See Table \ref{Table:VI-STV} and Fig. \ref{Fan-2016-ApJS-0235-LC-STV}(d).

 During the period of JD 2454706 to JD 2454785,  the light curve shows three peaks with R = 14.19, 14.2979, and 15.0057 with
  an interval between any close on two peaks being about 20 days. From JD 2454707 to JD 2454733, its brightness increased, and post JD 2454754, it decreased with two peaks. See Table \ref{Table:VI-STV} and Fig. \ref{Fan-2016-ApJS-0235-LC-STV}(e).

  At JD 2454730, we have 46 sets of data as shown in Fig. \ref{Fan-2016-ApJS-0235-LC-IDV}(c), it shows a variation of $\Delta R  \sim -0.11 $ mag over 113 min. corresponding to $A \,=10.87\%$, $\Delta T_v = 144.5 $ min, and  $C\,=\, 5.1$.
  At  JD 2454752, we have 37 sets of data as shown in Fig. \ref{Fan-2016-ApJS-0235-LC-IDV}(d), it shows a variation of $\Delta R  \sim 0.15 $ mag within 5.83 hrs corresponding to $A \,=\, 10.74\%$ and $\Delta t_v\, = \,499$ min. See Table \ref{Table:VI-IDV}

During the period of JD 2454797  to JD 22454800, its brightness dimmed from R = 16.1707 to R = 16.8523 corresponding to a $\Delta R  $ = 0.6816 mag over 3 days. In the period of JD 2455068 and 2455120, it dimmed from $R$ =  18.0956	to 18.4222 over 3 days, then it  brightened  to $R$ = 16.7378 within 22 days, and then dimmed to $R$ = 17.646  in 17 days. See Table \ref{Table:VI-STV} and Fig. \ref{Fan-2016-ApJS-0235-LC-STV}(f).

\newpage

\subsection{Radio Results}

At  radio band, we made observations during the period of Dec. 2006 to Nov. 2009, the results  are listed in Table \ref{Fan-2016-ApJS-0235-obs-rad} and shown in Fig. \ref{Fan-2016-ApJS-0235-LC-Radio}. We can see clearly that  during the optical observing period, radio band also shows two peaks:
peak 1 is at JD 2454267 with $f_{\rm peak} \, \sim$ 2.56 Jy and
peak 2 is at JD 2454779 with $f_{\rm peak} \, \sim$ 5.32 Jy. The corresponding interval is
 about 510 days.  From peak 2, we can see clearly that it brightened from $f \,\sim$  1.12 Jy to $f \,\sim$  5.32 Jy in 188 days and then dimmed to $f \,\sim$  1.24 Jy in 439 days showing a rapid rising and a slow decreasing.

\subsection{Discrete Correlation Function (DCF) Analysis}

Our long-term monitoring programs of AO 0235+164 were carried out with the 70-cm telescope at Abstumani Observatory, Georgia and the 25-m radio telescope at Xinjiang Astronomical Observatory, Chinese Academy of Sciences. We obtained a coverage of 12 years of optical data and   3 years of radio data. We hereby investigate whether there is any correlation and/or time delay between optical and radio bands.
To study this, we adopted the discrete correlation function (DCF) method to the
radio and optical data. The DCF method, which was described in details
\citep{Edelson1988} (also see Fan et al. 1998b), is intended for analysis
of the correlation of two data sets. This method can indicate the correlation
of two variable temporal series with a time lag, $\tau$.

Firstly, the set of unbinned correlation (UDCF) between data
points in the radio and optical data streams $a$ (for the optical data) and $b$
(for the radio data) is calculated by
\begin{eqnarray}
 {UDCF_{ij}}={\frac{ (a_{i}- \bar{a}) \times (b_{j}- \bar{b})}{\sqrt{\sigma_{a}^2 \times
 \sigma_{b}^2}}},
\label{UDCF}
\end{eqnarray}
where $a_{i}$ and $ b_{j}$ are points in the data sets, $\bar{a}$ and $\bar{b}$
 the averaged values of the data sets, $a_{i}$ and $ b_{j}$, and $\sigma_{a}$ and $\sigma_{b}$
the corresponding standard deviations. Secondly, averaging  the points
sharing the same time lag by binning the $UDCF_{ij}$ in the suitable sized
time-bins in order to get the $DCF$ for each time lag $\tau$:
\begin{eqnarray}
    {DCF(\tau)}=\frac{1}{M}\Sigma \;UDCF_{ij}(\tau), \label{DCF}
\end{eqnarray}
where $M$ is the total number of pairs. The standard deviation  for
each bin is
\begin{eqnarray}
    \sigma (\tau) =\frac{1}{M-1} \{ \Sigma\; [
    UDCF_{ij}-DCF(\tau) ]^{2} \}^{0.5}.  \label{sigma}
\end{eqnarray}

\begin{figure}
  \centering \includegraphics*[width=0.5\linewidth]{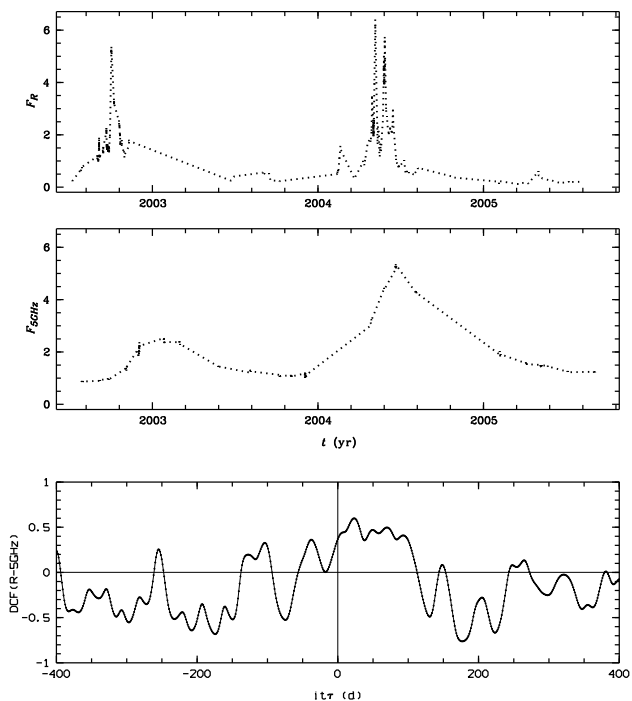}
  \caption{Optical  and radio light curves and DCF result for optical and radio light curves for AO 0235+164.
  Top panel is for optical light curve from our observations,
  middle panel is for radio light curve from our observations, and
  bottom panel is for the DCF result.}
\label{Fan-2016-ApJS-0235-RO-DCF}
\end{figure}

When relations (\ref{UDCF}) to (\ref{sigma}) are applied to the  optical R
flux density,  $F_{R} (\rm mJy)$, ($F_R\, (\rm mJy) = 3.08 \times 10^{6-0.4 m_R}$, Mead et al. 1990. $A_R$ = 0.173 is adopted for the R  magnitude) and the radio data, $f_{\rm GHz}$ (mJy),  and only the data carried out during the same observing period are considered,  a DCF
 result is obtained and shown in Fig.  \ref{Fan-2016-ApJS-0235-RO-DCF}. For  optical
and radio variation,  a marginal correlation is found with optical
variation leading radio variation by  23.2$\pm$12.9 days.

\subsection{Period Analysis}

\begin{figure}
    \centering \includegraphics*[width=0.6\linewidth]{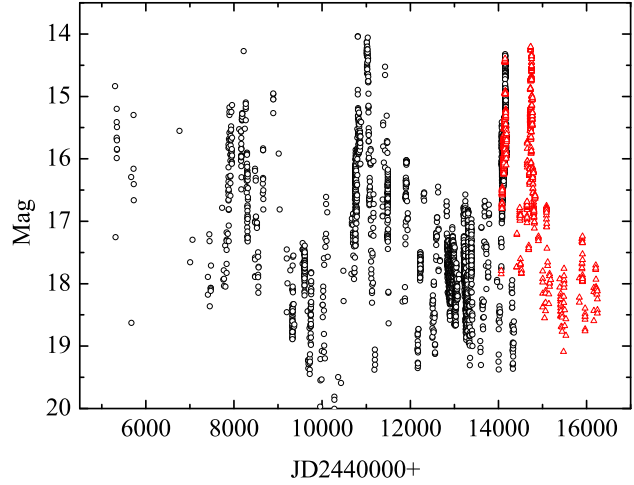}
    \caption{Historic light curve of AO 0235+164 during JD 2445300 to JD 2456247. The open circles are from the literatures and the triangles are from our own observations.}
  \label{Fan-2016-ApJS-0235-LC-whole}
  \end{figure}

 Now, we have compiled  the optical data from  literature \citep[also see][]{Fan2002,Wang2014}, and got an optical light curve covering a time span of 30 years   as shown in Fig. \ref{Fan-2016-ApJS-0235-LC-whole}. It is clear that the light curve is not evenly sampled,
 which makes periodicity analysis not easy. For the unevenly sampled time series, there are some periodicity analysis methods. Autocorrelation Function (ACF), Structure function (SF),  Jurkevich method and Power spectra density (PSD) for instance.

{\it Autocorrelation Function (ACF):}
{\it ACF} is the correlation of a time series with itself at different time. It is a time-domain tool for finding repeating patterns. In statistics, the {\it ACF} is the correlation between values of the time series at different times, as a function of the two times or of the time lag.

\begin{figure}
  \centering \includegraphics*[width=0.6\linewidth]{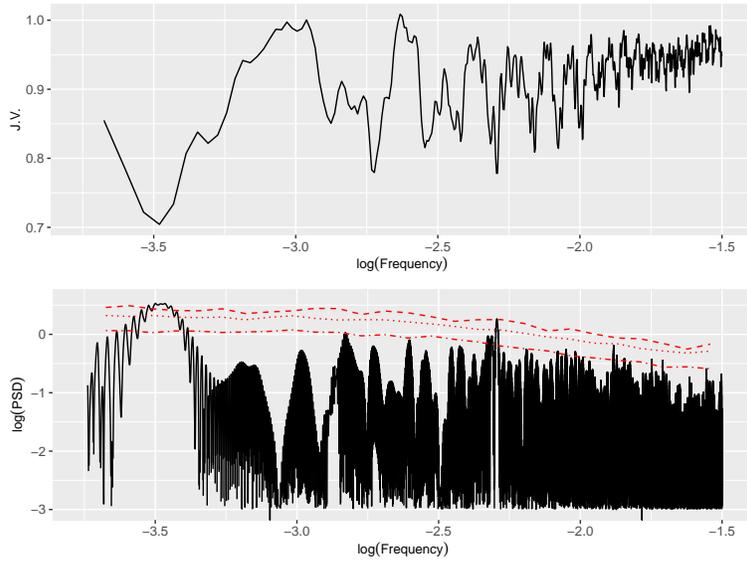}
  \caption{ Top panel: Periodicity analysis results of AO 0235+164 obtained by Jurkevich method. Bottom panel: results for AO 0235+164 by PSA method. The curves of the false alarm probability by using CAR1 method are also plotted. Two signals,
  $P_1$ = 8.26 yr, and $P_2$ = 0.54 yr, can be found with the threshold that $FAP \ge 3\sigma$.
  }
\label{Fan-2016-ApJS-0235-per}
\end{figure}

In the case of the unevenly sampled time series $X_{t_i},\,\,i=1,2\cdots N$,
the covariance between the time series $X_{t_i}$ and itself with a time lag $s$ is defined as
\begin{equation}  	
    \displaystyle C(s,m) = \frac{\sum_{i=1}^{N}\sum_{j=1}^{N}(X_{t_i}-\mu)(X_{t_j}-\mu)\,\exp{[-\frac{(t_i-t_j-s)^2}{2\,m^2}]} }{\sum_{i=1}^{N}\sum_{j=1}^{N}\exp{[-\frac{(t_i-t_j-s)^2}{2\,m^2}] }} ,
\end{equation}
where $\mu$ is the mean of $X_{t_i}$ and $m$ is the width of the gaussian weight function. As the covariance defined above, it can be noted that there are two variances $\sigma(s,m)_1$ and $\sigma(s,m)_2$, and the two variances could be defined as
\begin{equation}
	\displaystyle \sigma(s,m)_1 = \left(\frac{\sum_{i=1}^{N}\sum_{j=1}^{N}(X_{t_i}-\mu)^2\,\exp{[-\frac{(t_i-t_j-s)^2}{2\,m^2}]} }{\sum_{i=1}^{N}\sum_{j=1}^{N}\exp{[-\frac{(t_i-t_j-s)^2}{2\,m^2}] }}\right)^{-\frac{1}{2}},
	\end{equation}
	\begin{equation}
		\displaystyle \sigma(s,m)_2 = \left(\frac{\sum_{i=1}^{N}\sum_{j=1}^{N}(X_{t_j}-\mu)^2\,\exp{[-\frac{(t_i-t_j-s)^2}{2\,m^2}]} }{\sum_{i=1}^{N}\sum_{j=1}^{N}\exp{[-\frac{(t_i-t_j-s)^2}{2\,m^2}] }}\right)^{-\frac{1}{2}}.
		\end{equation}
Then the definition of the autocorrelation of $X_{t_i} $ is
    \begin{equation}\displaystyle R(s,m) = \frac{C(s,m)}{\sigma(s,m)_1\,m\sigma(s,m)_2}.\end{equation}
The {\it ACF} of $X_{t_i}$ is, itself, periodic with the same period. Periodicity analysis results of AO 0235+164 obtained by {\it ACF} method are shown in the top panel of Fig \ref{Fan-2016-ApJS-0235-per}.

{\it Structure Function (SF):}
Structure function were first considered by \citep{1941DoSSR..30..301K,1941DoSSR..32...16K}.
 The {\it SF} was discussed in papers \citep{1981ApPhy..24..323S,Emmanoulopoulos2010}. Similar
 to the {\it ACF}, the {\it SF} can also be a tool of periodic analysis.

For the unevenly sampled time series $X_{t_i}$, the {\it SF} of $X_{t_i}$ and itself with a time lag $s$ is defined as

\begin{equation}  	
    \displaystyle S(s,m) = \frac{\sum_{i=1}^{N}\sum_{j=1}^{N}(X_{t_i}^2-X_{t_j}^2)\,\exp{[-\frac{(t_i-t_j-s)^2}{2\,m^2}]} }{\sum_{i=1}^{N}\sum_{j=1}^{N}\exp{[-\frac{(t_i-t_j-s)^2}{2\,m^2}] }} .
\end{equation}

The {\it SF} of $X_{t_i}$ is, itself, peaked at the time point of period. Periodicity analysis results of AO 0235+164 obtained by {\it SF} method are also shown in the second panel from the top of Fig \ref{Fan-2016-ApJS-0235-per}.

{\it Jurkevich Method($JV$):}
 In the case that data are unevenly sampled in time series,  as we have done in \citep{Fan2014b}, the Jurkevich  method ({\it JV}) \citep{Jurkevich1971} and improved power spectral analysis (PSA) will be adopted for the possible periodicities.

 The {\it JV} is based on the expected mean square deviation. The deviation $V^2_m(\tau)$ of a giving period $\tau$ for a light curve $X(t_i),i=1,2,\cdots,N$ would be calculated, as described in \citep{Liu2011,Fan2014b}. If a frequency $f=1/\tau$ is equal to the true frequency, then $V^2_m(\tau)$ reaches the minimum. The plot of $V^2_m(f)$ against the $f$ is shown in the third panel from the top of Fig \ref{Fan-2016-ApJS-0235-per}.

{\it Power spectra density (PSD):} Many attempts of power spectral analysis have been made to investigate the periodicity.
An improved technique is the DCDFT+CLEANest ({\em DCDFT: Date-Compensated
  Discrete Fourier Transform}), \citep{Fer81,Fos95}, a least-square regression
on $\sin(\omega t)$, $\cos(\omega t)$ and constant function. The DCDFT is a
powerful method for unevenly spaced data, we adopt it to the light curve following Foster (1995).

In the case of unevenly sampled data, irregular spacing introduces myriad
complications into the Fourier transform. It will alter the peak frequency
(slightly) and amplitude (greatly), even introduce extremely large false peaks.
Following proposal by Foster (1995), we also used a CLEANest analysis to clean false
periodicities. The CLEANest algorithm can remove false peaks. Firstly, the
strongest single peak and corresponding false components are subtracted from the
original spectrum, then the residual spectrum is scanned to determine whether
the strongest remaining peak is still statistically significant. If so, then the
original data are analyzed to find the pairs of frequencies which best models
the data, these 2 peaks and corresponding false components are subtracted, and
the residual spectrum is scanned. By repeating the process, producing CLEANest
spectrum, till all statistically significant frequencies are included.

We assume that there are 7 independent frequency components to clean the
observation data, the CLEANest spectrum is shown in Fig \ref{Fan-2016-ApJS-0235-per}. Only two  periods:
$P_1$ = 8.26 yr and  $P_2$ = 0.54 yr can be found with the threshold that FAP $\geq \, 3\sigma$.

We also use the first order continuous autoregressive process (CAR1) to estimate the false alarm probability (FAP) of red noise. A CAR1 process $Y_t$, is a model of red noise, can be given by the stochastic difference function \citep{Brockwell2002}:

     $$Y_t = -\frac{1}{\tau}Y_t\,{\mathrm d}\,t+\,{\mathrm d}\,W_t.$$
Here, $\tau$ is the timescale of the CAR1 process, and $W_t$ is the Weiner process. The value of $\tau$ can be obtained from the light curve \citep{Mudelsee2002}, $\tau = 377^{+54}_{-17}$ days. The simulated $FAP$ curves are also shown in Fig \ref{Fan-2016-ApJS-0235-per}.  We can find only two possible periods, $P_1$ = 8.26 yr, and $P_2$ = 0.54 yr, which show that threshold $FAP \ge 3\sigma$.

\begin{figure}
  \centering \includegraphics*[width=0.6\linewidth]{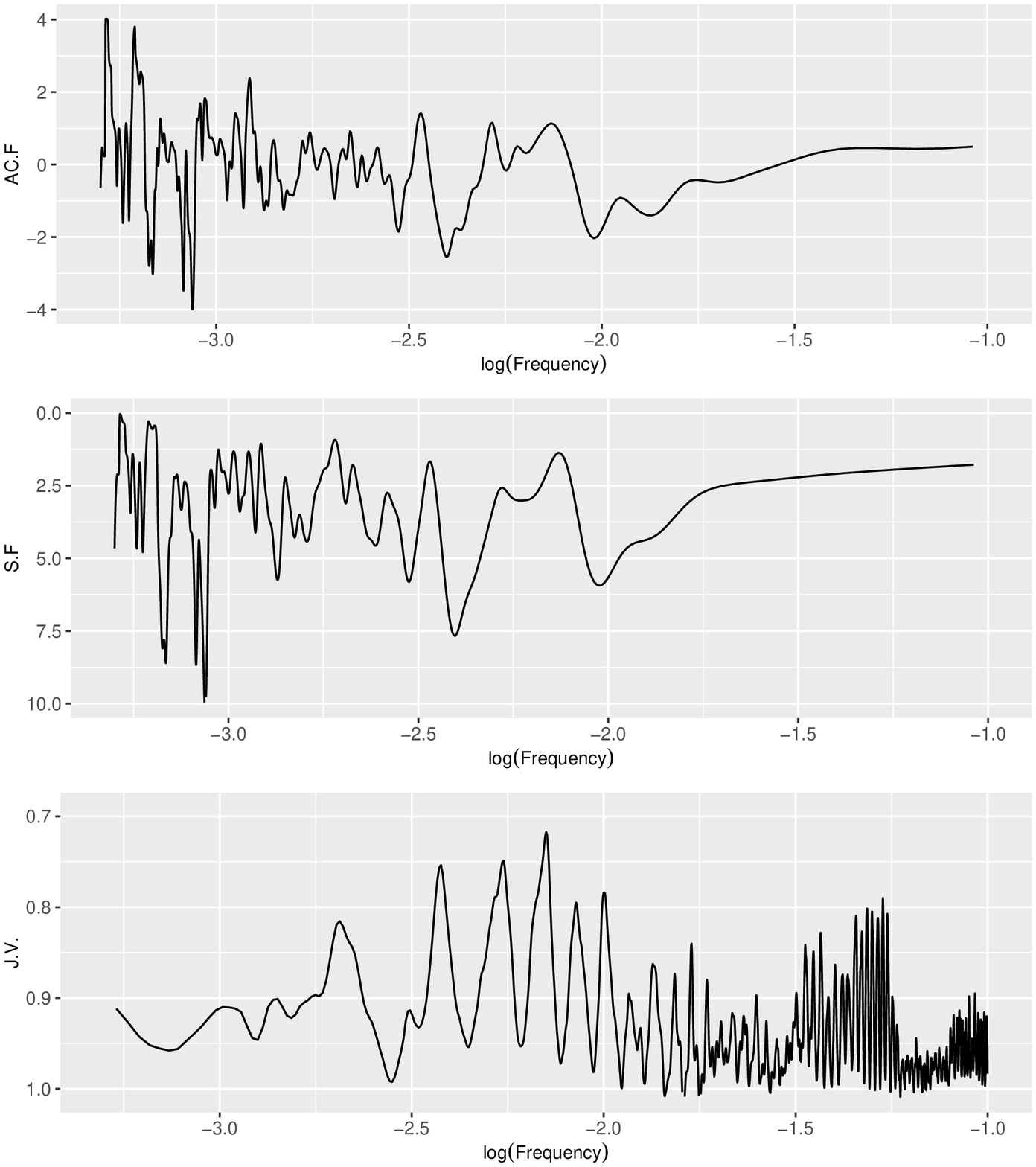}
  \caption{ Periodicity analysis results of AO 0235+164 based on our own observations and obtained by different methods:
  Top panel: Three signals of QPOs, $P_{1,ACF} = 0.37\pm 0.03 \,\mathrm{yr}$, $P_{2,ACF} = 0.52\pm0.02 \,\mathrm{yr}$ and $P_{3,ACF} = 0.77\pm0.04 \,\mathrm{yr}$ by {\it ACF} method;
Middle panel:  $P_{2,SF} = 0.37\pm 0.03 \,\mathrm{yr}$, $P_{2,SF} = 0.52\pm0.02 \,\mathrm{yr}$ and $P_{3,SF} = 0.77\pm0.04 \,\mathrm{yr}$  by {\it SF} method;
Bottom panel:  $P_{1,JV} = 0.39\pm 0.03 \,\mathrm{yr}$, $P_{2,ACF} = 0.50\pm0.02 \,\mathrm{yr}$ and $P_{3,ACF} = 0.72\pm0.05 \,\mathrm{yr}$ by {\it JV} method.
  }
\label{Fan-2016-ApJS-0235-opt-per}
\end{figure}

 When the methods are adopted to our optical data ( Nov, 2006 to Dec. 2012), the analysis results are shown in Fig. \ref{Fan-2016-ApJS-0235-opt-per}.
From the figure, we can see that there are three signals of QPOs, $P_{1,ACF} = 0.37\pm 0.03 \,\mathrm{yr}$, $P_{2,ACF} = 0.52\pm0.02 \,\mathrm{yr}$ and $P_{3,ACF} = 0.77\pm0.04 \,\mathrm{yr}$ can be derived by using {\it ACF} method, three signals of QPOs, $P_{2,SF} = 0.37\pm 0.03 \,\mathrm{yr}$, $P_{2,SF} = 0.52\pm0.02 \,\mathrm{yr}$ and $P_{3,SF} = 0.77\pm0.04 \,\mathrm{yr}$ can be found by using {\it SF} method, and three similar signals of QPOs, $P_{1,JV} = 0.39\pm 0.03 \,\mathrm{yr}$, $P_{2,ACF} = 0.50\pm0.02 \,\mathrm{yr}$ and $P_{3,ACF} = 0.72\pm0.05 \,\mathrm{yr}$ can be obtained by using {\it JV} method. Based on the above results, there are QPOs but not identical but nearly similar period, their  average value gives, $P_{avg} = 0.55\pm 0.03 \,\mathrm{yr}$. The $P_{avg}$ is consistent with the 0.54 yr period sign.

\section{Discussions and Conclusions}

Variability is one of the extreme properties of BL Lacertae objects. Variability can be divided into three classes according to their time scales:
 micro-variability with time scale being less than one day;
 short-term variability with time scale being a few days to months;
 long-term variation with time scale being years \citep{Fan05}. But it is not easy to say whether a variation is real or not.  It can be taken as a real variability if the variability shows up in simultaneous multiwavelength observations. Unfortunately, it is not always the case for one to do simultaneous multiwavelength observations, and most of monitoring programs are performed at a certain waveband.
 Romero et al. (1999) introduced a variability parameter $C$, to justify a variation to be strong or not. de Diego (2010) introduced a $F$-test for the discussion of variability (see also Gaur et al. 2012).
 Heidt \& Wagner (1996) proposed a method to obtain a variability amplitude $A$. We also proposed that  a variation can be taken as a real one  if the variability is 3 times greater than the deviation, namely $\Delta m_{12}= m_{1} - m_{2} \ge 3 \sqrt{\sigma_1^2+\sigma_2^2}$, where $\sigma_1$ and $\sigma_2$ are the uncertainties corresponding to m$_{1}$ and m$_{2}$, the corresponding time interval is adopted as the time scale $\Delta$ T = t$_{m2}$-t$_{m1}$ \citep{Fan2009}. Time scale is also defined to be $\Delta T_v = \Delta F /(dF/dt)$.

AO 0235 + 164 is one of the well studied  objects. Although it was classified as a BL Lac object, the equivalent width (EW) of emission
 lines  varies from one observational epoch to another. It is violently variable in all wavebands from radio through high energetic $\gamma$-rays \citep{Ackermann2012}. Its synchrotron peak frequencies obtained by different authors are
$\rm log \nu_p\,(Hz)$ =
13.29 \citep{Fan2016a},
13.39 \citep{Sambruna1996},
13.57 \citep{Nieppola2006}, and
13.10 \citep{Abdo2010},
suggesting it to be a low-synchrotron peaked (LSP) BL Lac. During the  2008 September to 2009
February outburst  period, the $\gamma$-ray activity is correlated with near infrared/optical flares, but there is no clear correlation between  $\gamma$-ray and radio emissions \citep{Ackermann2012}.

Intraday radio variabilities were detected in the works  \citep{Quirrenbach1992,Romero1997,Kraus1999}. Its  unusual radio outburst detected at $\lambda \, =20 \rm {cm}$ results in  a brightness temperature of $T_{\rm B} \sim 7 \times 10^{17\circ}  {\rm K}$ for $\Delta t_{20\, \rm{cm}} \sim 2 \rm {days}$ \citep{Kraus1999}. Such a high brightness temperature suggests
 a very large  Doppler factor of $\delta \, \sim 100$. For the source, radio components show superluminal velocities  as fast as $\beta \sim 30 h^{-1}$ ($h = H/100 {\rm km \, Mpc^{-1}\, s^{-1}}$) \citep{Abraham1993,Chu1996, Jorstad2001}.  Let $h \sim 0.67$, $\beta \sim 30 h^{-1}$ suggests a $\delta\,\, \sim\, 2\Gamma \,\, = 2\, \beta \sim 90$.

At the optical bands, rapid variability was reported by many authors \citep{Schramm1994,Heidt1996,Romero2000,Gaur2012}.
Schramm et al.  (1994) reported an extreme optical variability of 1.6 mag within 48 hours.
Heidt \& Wagner (1996) found variation amplitude of 6.33\% per day.
Romero et al. (2000) reported an intra-night variability with amplitudes of $A\,\sim 100\%$ over 24 hours, variations of $\Delta\, m = 0.5$ mag were detected in R and V bands  within a single night, and variations up to 1.2 magnitudes occurred from night to night.

In the soft X-ray region, ROSAT detected an increasing by a  factor of 1.7 in about 3 days, and a decreasing of a  factor of 3.5 in about 13 days \citep{Urry1996}, which suggest doubling time scales of $\Delta\, T_D = 1.76$ and 3.71 days respectively.

\subsection{Variability}

For AO 0235+164, its historic  variation amplitude is as large as $\Delta \, m \, \sim 5.0$ mag \citep{Rieke1976,Stein1976,FanL2000}. In our monitoring period, the light curve shows a variation amplitude of $\Delta\, R \, \sim \, 4.88$ mag and 6 peaks, the 6 peaks show intervals of 1.65, 0.93, 1.08, 1.15, and 0.82 years. The largest variation amplitude in our observations is similar to the historically largest amplitude.

It shows different time scales,
 a brightness decreasing of $\Delta R$ = 0.8173 mag over 12 days (JD 2454168 to JD 2454180),
  $\Delta R = 0.6816$ mag over 3 days (JD 2454797 to JD 22454800) and 0.3366 mag over 3 days (JD 2455068 and 2455120),
  $\Delta R$ = 0.528 mag  over 2 days (JD 2454140-JD 2454142),
 and $\Delta R$ = 0.627 mag ($A = 62.44\%$) over 1 day (JD 2454124 to JD 2454125).
 A brightness increase of $\Delta R$ = 0.4204 mag over 10 days (JD 2454180 to JD 2454190),
  $\Delta R$ = 1.3037 mag over 7 days (JD 2454649 to JD 2454656),
  $\Delta R$ = 1.537 mag within 5 days (JD 2454146 to JD 2454151), and
  $\Delta R = 0.7367$ mag ($A = 73.35\%$) over 2 days (JD 2454122 to JD 2454124).  A variation over 2 days was found in a paper by Sagar et al. (2004), who noticed a dramatic brightness fading of 0.83 mag within 2 days in 1999.

  Very rapid variabilities are also detected as listed in Table  \ref{Table:VI-IDV}. Our monitoring
 results show that the shortest time scale is $\Delta\, T_v$ = 73.5 min, which is longer than  the time scale of 0.31 hrs detected in the polarization variability in V band \citep{Hagen2008}.

\subsection{Correlation between Optical and Radio Bands}

The optical/radio correlation is investigated for
 AO 0235 + 164 (see Takalo et al. 1992 and references therein).
A positive correlation with a delay of $0\, \sim \,2$ months from optical to radio was reported in a paper \citep{Clements1995}. From our DCF calculation based on our own optical and radio data, we found that there is a marginal correlation with optical
 variation leading  radio variation by  23.2$\pm$12.9 days. Our results are consistent with the time-lag of $0\, \sim \,2$ months
by Clements et al. (1995).

  AO 0235 + 164 has a complex feature, it shows absorption and emission lines corresponding to
redshifts of 0.542 and 0.851 (Burbidge et al. 1976, Smith et al. 1985),  0.92 (Cohen et al. 1987), and 0.94 (Charles, et al. 1994).
AO 0235+164 is one of the candidates for the discussions of microlesing phenomena
(see
Charles et al. 1994,
Webb et al. 2000,
Rain et al. 2009,
Gaur2012
).
If the outbursts are due to microlensing, the simplest scenarios
predict symmetric outbursts that are frequency independent.  In our observations, the source display sharp outbursts in its optical band in 2002 and 2004 while  broad and delayed outbursts in the radio band at the end of 2002/beginning of 2003 and in the middle of 2004. The outbursts in the optical band in 2004 have double peaks while that in the radio band is only one broad peak. The DCF analysis shows that the optical variability leads the radio variability by  23.2$\pm$12.9 days. So, the outbursts in the optical and radio bands maybe not caused by the microlensing effect.

\subsection{Periods}

Periodicity analysis is also interesting in active galactic nuclei (AGNs). It was claimed to show up in 3C 120 \citep{Jurkevich1971}, then a lot of AGNs have been reported to show quasi-periodicity signs in their light curves \citep{Sillanpaa1988,Fan1998,Fan2002,Fan2007,Fan2010,Fan2014b,Cia04,Wu2006, Ciprini07,Valtonen2008, Webb2000,Rani2010,QianT2004,Gupta2014,Wang2014,Bon2016,Wang2016}.  As one of the well studied blazars, AO 0235+164 has been observed and periodicity analyzed in many works \citep{Webb1988,Smith1995,Fan2002,Fan2007,Raiteri2001,Raiteri2006,Raiteri2007,Raiteri2008,Wang2014}.

Using discrete Fourier transform(DFT), Webb et al. (1988) analyzed the optical light curve of AO 0235+164 and found periods of
2.79, 1.53 and 1.29 years, later on they removed a linear trend from the light curve and adopted unequal-interval Fourier transform and CLEAN techniques to its normalized data, and obtained periods of 2.7 and 1.2 years.
Smith \& Nair (1995) found periods of 2.7 and 3.6 years in the optical band.
Raiteri et al.(2001) reported a 5.67 year period in  the R optical light curve and  1.8, 2.8 and 3.7 years in radio bands. In our previous papers \citep{Fan2002, Fan2007}, we found periods of 2.0 years,  2.95$\pm$0.15 years and 5.87$\pm$1.3 years in the optical light curve, and period of 10.0$\pm$1.3 years, 5.7$\pm$0.3 years and 5.8$\pm$0.3 years at 4.8GHz, 8.0 GHz, and 14.5 GHZ light curves.

Raiteri et al. (2001) claimed a possible quasi-periodic occurrence of the major radio and optical outbursts of AO
0235+164 every 5.7$\pm$0.5 years, which result in the  WEBT campaign observing the source \citep{Raiteri2006,Raiteri2007}. The period analysis based on the historic and the campaign observations suggests that the period for the large outburst is 8 $\sim$ 8.5 years \citep{Raiteri2006,Raiteri2007, Raiteri2008,Gup08}. It is clear that the periodicity analysis results depend on the observations, that is why different works give different results.

In this work, we find that two periods $P_1$ = 8.26 yr and $P_2$ = 0.54$\sim$ 0.56 yr have $FAP \geq 3\sigma$  in the optical light curve.
However, the time coverage used to investigate the period is only 30 years.
As mentioned by Koen (1990)  that estimating (rather than knowing) the variance of the time series can change the $FAP$s by orders of magnitude. Therefore, the periods should be investigated using more dense observations.

\subsection{Conclusions}

In this work, we have presented the R optical observations during the period of Nov, 2006 to Dec. 2012  and  radio observations light curve during the period of Dec. 2006 to Nov. 2009. Following results are obtained:

1. Largest variation is $\Delta R $ = 4.88 mag from our observations.

2. The shortest time scale detected in our monitoring program is $\Delta T_v$ = 73.5 min.

3. The optical and radio variation is correlated with optical leading radio by  23.2$\pm$12.9 days.

\section*{acknowledgements}

The work is partially supported by the
National Natural Science Foundation of China (NSFC U1531245, U1431112, U11203007, 11403006, 10633010, 11173009), the Innovation Foundation of Guangzhou University (IFGZ),
Guangdong Province Universities and Colleges Pearl River Scholar
Funded Scheme(GDUPS)(2009), Guangdong Innovation Team for Astrophysics(2014KCXTD014), Yangcheng Scholar Funded
Scheme(10A027S), and support for Astrophysics  Key Subjects of Guangdong Province and Guangzhou City. The Abastumani team acknowledges financial support of the project FR/639/6-320/12 by the Shota Rustaveli National Science Foundation under contract 31/76.

\begin{deluxetable}{ccc}
  \tabletypesize{\scriptsize} \tablecaption{Observational data for
    AO 0235 + 164} \tablewidth{0pt}
    \tablehead{
    \colhead{JD} & \colhead{R-Mag.} &  \colhead{$\sigma$}  \\
    (1) & (2) & (3) } \startdata
2454065.44811 	&	17.8296	&	0.0561	\\
2454065.45163 	&	17.7653	&	0.1019	\\
2454082.33861 	&	16.7107	&	0.0247	\\
2454082.34072 	&	16.6765	&	0.0254	\\
2454082.34281 	&	16.6997	&	0.025	\\
2454082.34492 	&	16.728	&	0.0261	\\
2454082.34703 	&	16.7373	&	0.0255	\\
2454082.34912 	&	16.7274	&	0.0248	\\
\enddata
  \label{Fan-2016-ApJS-0235-obs-opt}
\end{deluxetable}

\begin{deluxetable}{ccc}
  \tabletypesize{\scriptsize} \tablecaption{Radio Observations  for
    AO 0235 + 164} \tablewidth{0pt}
    \tablehead{
    \colhead{JD} & \colhead{F$_{4.8GHz}$-Jy} &  \colhead{$\sigma$}  \\
    (1) & (2) & (3) } \startdata
2454085.6533	&	0.8753	&	0.0078	\\
2454085.6885	&	0.8744	&	0.0074	\\
2454085.7237	&	0.8755	&	0.0075	\\
2454085.7571	&	0.8746	&	0.0075	\\
2454085.7901	&	0.8750	&	0.0075	\\
2454085.8269	&	0.8897	&	0.0077	\\
2454085.8608	&	0.8744	&	0.0078	\\
2454086.5987	&	0.8764	&	0.0074	\\
\enddata
  \label{Fan-2016-ApJS-0235-obs-rad}
\end{deluxetable}

\end{document}